\documentclass[lettersize,journal]{IEEEtran}
\usepackage{amsmath,amsfonts}
\usepackage{algorithmic}
\usepackage{array}
\usepackage[caption=false,font=normalsize,labelfont=sf,textfont=sf]{subfig}
\usepackage{textcomp}
\usepackage{stfloats}
\usepackage{url}
\usepackage{verbatim}
\usepackage{graphicx}
\usepackage{balance}
\usepackage{xcolor}
\usepackage{epstopdf}
\usepackage{multirow}
\usepackage{booktabs}
\usepackage{latexsym}
\usepackage{cite}
\usepackage{algorithm}
\usepackage{hyperref}
\usepackage{marvosym}
\usepackage[numbers, sort&compress]{natbib}
\usepackage{nomencl}
\usepackage{textcomp,mathcomp}
\usepackage{color}
\usepackage{enumitem}

\newenvironment{termtable}[1][2cm]{%
	\def\term##1##2{\item[$##1$] ##2}%
	\itemize[left=0pt .. #1, itemindent=0pt,
	align=parleft, nosep]
}
{%
	\enditemize
}
\makenomenclature

\begin{document}
	
	\title{Multi-Objective Sizing Optimization Method of Microgrid Considering Cost and Carbon Emissions}
	
	\author{Xiang Zhu,~\IEEEmembership{Graduate Student Member,~IEEE,}
		Guangchun Ruan,~\IEEEmembership{Member,~IEEE,}
		Hua Geng,~\IEEEmembership{Fellow,~IEEE,}\\
		Honghai Liu,
		Mingfei Bai,
		and~Chao Peng
		\thanks{\emph{(Correspondence author: Hua Geng)}\\
			X. Zhu and H. Geng are with the Department of Automation, Tsinghua University, Beijing 10084, China, and Beijing National Research Center for Information Science and Technology, Tsinghua University, Beijing 10084, China (e-mail: zhu-x22@mails.tsinghua.edu.cn; genghua@tsinghua.edu.cn).}
		\thanks{G. Ruan is with the Laboratory for Information \& Decision Systems (LIDS), MIT, Boston, MA 02139 (email: gruan@mit.edu).}
		\thanks{H. Liu and M. Bai are with the State Key Laboratory of Alternate Electrical Power System with Renewable Energy Sources and School of Electrical \& Electronic Engineering, North China Electric Power University, Beijing, 102206, China (email: hbdl\_liuhonghai@163.com; mingfeibai\_tut@126.com).}			
		\thanks{C. Peng is with the School of Automation Engineering, University of Electronics and Technology of China, Chengdu 611731, China (email: pcddiy@163.com).}
		
	}
	
	\maketitle
	
	\begin{abstract}
		Microgrid serves as a promising solution to integrate and manage distributed renewable energy resources. In this paper, we establish a stochastic multi-objective sizing optimization (SMOSO) model for microgrid planning, which fully captures the battery degradation characteristics and the total carbon emissions. The microgrid operator aims to simultaneously maximize the economic benefits and minimize carbon emissions, and the degradation of the battery energy storage system (BESS) is modeled as a nonlinear function of power throughput. A self-adaptive multi-objective genetic algorithm (SAMOGA) is proposed to solve the SMOSO model, and this algorithm is enhanced by pre-grouped hierarchical selection and self-adaptive probabilities of crossover and mutation. Several case studies are conducted to determine the microgrid size by analyzing Pareto frontiers, and the simulation results validate that the proposed method has superior performance over other algorithms on the solution quality of optimum and diversity.  
	\end{abstract}

	\begin{IEEEkeywords}
		distributed energy resource, low carbon, battery energy storage system, renewable energy, stochastic multi-objective optimization, genetic algorithm
	\end{IEEEkeywords}
	
	\IEEEpeerreviewmaketitle
	\section*{Nomenclature}
	
	\subsection*{\bf{Abbreviations}}
	\begin{termtable}[1.9cm]
		\term{PV}{Photovoltaic}
		\term{WT}{Wind turbine}
		\term{DG}{Diesel generator}
		\term{PEC}{Pollutant emissions by $CO_{2}$}
		\term{DER}{Distributed energy resource}
		\term{BESS}{Battery energy storage system}
		\term{LPSP}{The loss of power supply probability}
		\term{SMOSO}{Stochastic multi-objective sizing optimization}
		\term{SAMOGA}{Self-adaptive multi-objective genetic algorithm}
	\end{termtable}
	
	\subsection*{\bf{Sets}}
	\begin{termtable}[1.9cm]
		\term{T}{Set of time periods}
		\term{W}{Set of scenarios}
	\end{termtable}
	
	\subsection*{\bf{Parameters}}
	\begin{termtable}[1.9cm]
		\term{N_{max}^{WT}}{Maximum number of WT}
		\term{N_{max}^{PV}}{Maximum number of PV}
		\term{N_{max}^{DG}}{Maximum number of DG}
		\term{N_{max}^{ES}}{Maximum number of batteries}
		\term{\eta_{Lmax}}{Maximum allowable LPSP [\%]}
		\term{P^{W}_{i,t}}{Power generation of WT $i$ at time period $t$ [kW]}
		\term{P^{P}_{i,t}}{Power generation of PV $i$ at time period $t$ [kW]}
		\term{P^{LD}_{t}}{Power load at time period $t$ [kW]}
		
		\term{\underline{E}^{ES}_{i},\overline{E}^{ES}_{i}}{Lower/Upper bound of energy of battery $i$ [kWh]}
		\term{\eta^{ch}_{i},\eta^{dc}_{i}}{Charging/discharging efficiency of battery $i$ [\%]}
		\term{R^{D+}_{i},R^{D-}_{i}}{Upward/Downward ramping limits of DG $i$ [kW]}
		\term{R^{SU}_{i},R^{SD}_{i}}{Start-up/Shut-down ramping limits of DG $i$ [kW]}
		\term{\pi_{w}}{Probability of stochastic scenario $w$ occurrence}
		\term{\underline{P}^{D}_{i},\overline{P}^{D}_{i}}{Lower/Upper bound of generation of DG $i$ [kW]}
		\term{P_{c},P_{m}}{The adaptive probabilities of cross and mutation}
		\term{P_{c,0},P_{m,0}}{The initial value of probabilities of cross and mutation}
		\term{Q_{i}^{max}}{The end-of-life retained capacity percentage [\%]}
		\term{T_{en}}{The environmental temperature [K]}
	\end{termtable}
	
	\subsection*{\bf{Variables}}
	\begin{termtable}[1.9cm]
		\term{N^{WT},N^{PV}}{Number of four types of DERs in microgrid}
		\term{N^{DG},N^{ES}}{}
		\term{N^{re}}{The total number of batteries need to be replaced}
		\term{e^{ES}_{i,t}}{Energy state of battery $i$ at time period $t$ [kWh]}
		\term{p^{ch}_{i,t},p^{dc}_{i,t}}{Charging/discharging power of battery $i$ at time period $t$ [kW]}
		\term{Q_{i,t}}{Capacity loss of battery $i$ at time period $t$ [\%]}
		\term{u_{i,t}}{Binary variables denoting the on/off status of DG at time period $t$}
		\term{v^{SU}_{i,t},v^{SD}_{i,t}}{Binary variables denoting the start-up/shut-down status of DG at time period $t$}
		\term{P^{D}_{i,t}}{Generation of DG $i$ at time period $t$ [kW]}
		\term{P^{G}_{t}}{Power traded at time period $t$ [kW]}
	\end{termtable}

	\section{Introduction}
	
	\IEEEPARstart{T}HERE is a global consensus that the future power grids will be dominated by renewable energy, especially through a promising form of distributed renewable energy resources. Microgrids may serve as a popular option to boost the utilization of these distributed resources and explore the potential benefits of emission reduction~\cite{jiayi2008review,borges2012overview,jiang2018coordinated}. Considering the natural intermittency of renewable energy resources (RES), diesel generator (DG) and battery energy storage system (BESS) are deployed in microgrids as auxiliary power generation modes \cite{jia2019distributed,rezkallah2019comprehensive,dehnavi2018distributed,zhu2022multi}. Modern microgrids have prominent uncertainty and complex composition, which bring difficulties to the sizing optimization before its utilization. Therefore, uncertainty handling and multi-objective consideration are indispensable when determining the optimal sizing of microgrids \cite{cao2020hybrid}. 
	
	Stochastic optimization~(SO) is one of the mainstream methods for uncertainty modeling with high computational efficiency and the authors of \cite{SO-1} captured the uncertainty of RES and other factors within this framework.	In \cite{SO-2}, a two-stage stochastic model was established as a mixed-integer quadratic programming (MIQP). In \cite{SO-3}, the authors combined stochastic dynamic programming (SDP) with stochastic dual dynamic programming (SDDP) to address microgrid system energy management. Reference~\cite{SO-4} focused on uncertain islanding events that have multi-occurrence and multi-period characteristics and used Benders decomposition for solution. Reference~\cite{SO-5} applied the two-stage stochastic programming (TSSP) model to optimize the total costs of multi-microgrids under several uncertainty factors. Generally speaking, SO is widely used in microgrid sizing optimization or operation strategy to handle multiple uncertainties.
	
	Apart from dealing with uncertainty, microgrids need to account for a range of objectives because total cost minimization might not be the sole objective for operators when the system and power device scale increase~\cite{li2017multiobjective}. Therefore, multi-objective optimization (MOO) of microgrids is employed for the solution. Reference \cite{MOO-2} presented two MOO models to evaluate the scale of investment required in BESS sizing of the microgrid. With greater concern about climate change due to carbon emissions, i.e., $CO_{2}$, the novel combination of optimal objectives involving low carbon and the comprehensive cost was analyzed in \cite{CO2-1,CO2-2}. However, the above SO and MOO models ignore the degradation effects of BESS involved in microgrids, which will limit the accuracy and practicality of the solution.  
	
	BESS is an important component of microgrid energy management to improve the renewable energy absorption capacity as well as enhance the power stability of the system. It is important to consider the accurate evaluation of BESS operation via its real-life performance involving degradation effects \cite{BESS-1,BESS-7}. In \cite{BESS-2}, BESS degradation effects were reflected in the economic cost through linear modeling. Besides, linearization methods such as piecewise linear approximation \cite{BESS-5} and binomial expression approximation \cite{BESS-6} were deployed to solve the complex problem. Based on linear modeling or linearization method, the nonlinear problems can be reformulated as the Mixed Integer Linear Programming (MILP) framework, which can be solved using commercial solver CPLEX or Gurobi directly and effectively \cite{BESS-3}.	However, MILP is only suitable for linear or linearized battery degradation functions, which makes it difficult to accurately describe the nonlinear characteristics of degradation effects in real industrial applications \cite{degradation-2, degradation-3}.
	
	Therefore, the challenge is how to analytically evaluate the impact of the BESS nonlinear degradation effect on the sizing optimization of the microgrid without linearized approaches. Motivated by \cite{degradation-3, degradation-new1, degradation-new2, degradation-new3}, one of the main nonlinear degradation characteristics of BESS is the time-varying capacity due to the cumulative power throughput. The time-varying capacity of BESS changes the fundamental nature of microgrid sizing optimization because the traditional models consider the time-invariant capacity. Furthermore, this feature will transform the microgrid MOO model into a mixed integer nonlinear programming problem (MINLP), which makes the traditional solver-based approaches unable to be applied.

	To address the computational challenge, the genetic algorithm (GA), which is one effective heuristic algorithm to handle MINLP problems \cite{MOGA-1}, is employed to solve the nonlinear multi-objective optimization problem. The performance of GA largely depends on the probabilities of crossover and mutation operations. Considering the constant probabilities of crossover and mutation would limit the performance of solution quality due to the incapability of adapting changes during evolution, several adaptive genetic algorithms were proposed \cite{AGA-1, AGA-2, AGA-3}. However, the available methods rarely pay attention to the degree of population convergence, which will deteriorate the population diversity and search efficiency~\cite{ooi2019self}.

	To fill the research gaps mentioned above, this paper addresses the stochastic multi-objective sizing optimization of the grid-connected microgrid that involves uncertainties of RES as well as load demand and the nonlinear capacity reduction of BESS due to cumulative power throughput. The objective of the microgrid is to minimize the system cost and carbon emissions simultaneously. A new self-adaptive multi-objective genetic algorithm that fully considers the degree of population convergence is proposed methodologically.
	
	The main contributions are twofold:
	\begin{enumerate}
		\item A new stochastic multi-objective sizing optimization (SMOSO) model considering multiple uncertainties and nonlinear BESS degradation characteristics is established for maximizing the economy and minimizing carbon emissions simultaneously. The degradation effect of BESS is modeled as a nonlinear function of power throughput and reflected in time-varying capacity and corresponding economic cost. 
		
		\item A new self-adaptive multi-objective genetic algorithm (SAMOGA) based on the pre-grouped hierarchical selection, self-adaptive probabilities of crossover and mutation is proposed to effectively solve the SMOSO model. The proposed algorithm can well handle the nonlinear SMOSO problem and the modified crossover and mutation probabilities are changed adaptively according to the degree of convergence.
	\end{enumerate}
	
	The rest of this paper is organized as follows. Section~II presents the formulation of a grid-connected microgrid system. Section~III derives the scenario-based stochastic multi-objective sizing optimization model. Section~IV proposes the self-adaptive multi-objective genetic algorithm for solving the optimization model. Case studies are presented in Section~V. Finally, Section~VI concludes the paper.

	\section{Microgrid System}
	
\begin{figure}
		\centering
		\includegraphics[width=0.95\linewidth]{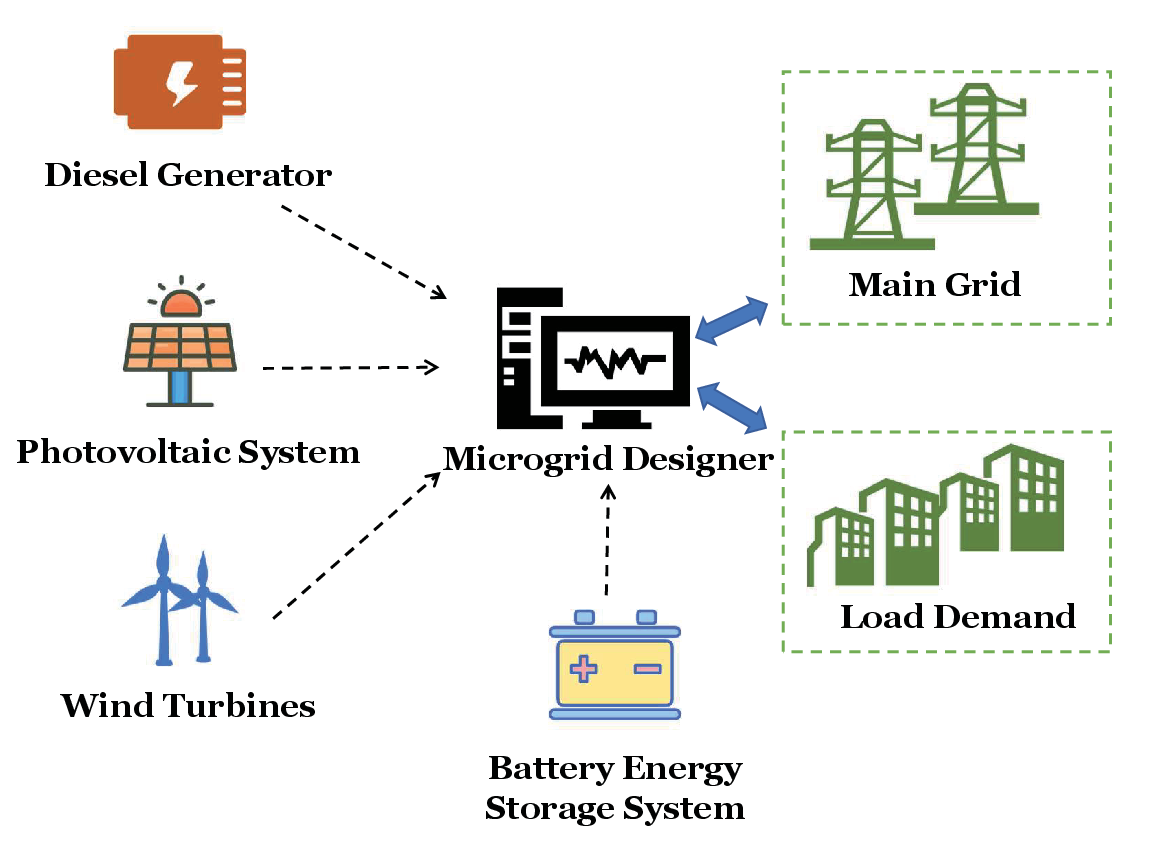}
		\caption{System framework of a typical microgrid.}
		\label{fig.demo}
	\end{figure}

	Fig.~\ref{fig.demo} gives an overview of the research system architecture. Here the system is a grid-connected microgrid with four types of distributed generators, i.e., wind turbines, solar panels, diesel generators, and battery storage devices. End-users and the main grid act as the power demand of this microgrid. More modeling details will next be discussed.
	
	\subsection{Modeling of Wind Turbines}
	
	The output power of wind turbine $i$ at time period $t$ can be calculated as follows \cite{peng2014research}.
	\begin{equation}
		P_{i,t}^{W} = \left\{
		\begin{aligned}
			&0      &,v_{t} \le v_{cin}
			\\
			&P^{WR}\frac{v_{t}^{3}-v_{cin}^{3}}{v_{r}^{3}-v_{cin}^{3}}&,v_{cin} < v_{t} \le v_{r}
			\\
			&P^{WR} &,v_{r} < v_{t} < v_{cout}
			\\
			&0      &,v_{t}\ge v_{cout}
		\end{aligned}
		\right.
		\label{eq.1}
	\end{equation}
	where $P^{WR}$ is the rated output power of a wind turbine; $v_{t}$ is the wind speed at time period $t$; $v_{cin}$, $v_{cout}$ and $v_{r}$ are the cut-in wind speed, cut-out wind speed and rated wind speed respectively.
	
	\subsection{Modeling of Photovoltaic~(PV) Panels}
	
	The output power of a PV device at time period $t$ can be calculated as follows \cite{peng2014research}.
	
	\begin{equation}
		P_{i,t}^{P} = P^{PR}\frac{G_{en}}{G_{stc}}[1+k_{p}(T_{en}-T_{stc})]
		\label{eq.2}
	\end{equation}
	where $P^{PR}$ is the rated output power of a PV device; $G_{en}$ and $T_{en}$ are the environmental illumination intensity and temperature; $G_{stc}$ and $T_{stc}$ are the reference values of the environmental illumination intensity and temperature respectively; $k_{p}$ is the temperature conversion coefficient.
	
	\subsection{Modeling of Battery Storage}
	\subsubsection{Conventional Model}
	The energy conversion model of the batteries can be represented as follows \cite{Kmeans}.
	
	\begin{equation}
		\begin{aligned}
			&e^{ES}_{i,t} = e^{ES}_{i,t-1}+\eta^{ch}_{i}p^{ch}_{i,t}-p^{dc}_{i,t}/\eta^{dc}_{i}, \forall t\in T, \forall i\in N^{ES}
			\\
			&0 \le p^{ch}_{i,t} \le \overline{P^{ch}_{i}},\forall t\in T, \forall i\in N^{ES}
			\\
			&0 \le p^{dc}_{i,t} \le \overline{P^{dc}_{i}},\forall t\in T, \forall i\in N^{ES}
			\\
			&\underline{E}^{ES}_{i} \le e^{ES}_{i,t} \le \overline{E}^{ES}_{i}, \forall t\in T, \forall i\in N^{ES}
		\end{aligned}
		\label{SOC-1}
	\end{equation}
	where $e^{ES}_{i,t}$ is the energy state of the battery $i$ at time period $t$; $p^{ch}_{i,t}$ and $p^{dc}_{i,t}$ are the charging power and discharging power of storage battery $i$ at time period $t$ respectively; $\eta^{ch}_{i}$ and $\eta^{dc}_{i}$ are the charging efficiency and discharging efficiency of battery $i$ respectively.
	
	It can be seen from the energy conversion model (\ref{SOC-1}) that the capacity $\overline{E}^{ES}_{i}$ of battery~$i$ is constant, which will be modified in the following part.

	\subsubsection{Degradation Model}
	In the conventional model of storage batteries, the degradation effect of the battery is ignored, which will cause the model of BESS not to be accurate enough to analyze the reliability of the solution in terms of delivered energy and cost. As for the sizing optimization of the microgrid, the change of capacity of BESS due to cumulative loss has a significant impact, which is determined by the power throughput and reflected in time-varying capacity corresponding economic cost. Derived from credible physical experiments, the formulations of capacity change are as follows \cite{degradation-3, degradation-new1, degradation-new2, degradation-new3}.
	
	\begin{equation}\label{degra-2}
		Q_{i,t} = \kappa \cdot exp ({\frac{Ea}{R \cdot T_{en}}}) \cdot (AH_{i,t})^{z},\forall i\in N^{ES}, \forall t\in T
	\end{equation}
	where 
	\begin{equation}\label{degra-3}
		AH_{i,t} = \sum_{\tau = 1}^{t}(p^{dc}_{i,\tau}+p^{ch}_{i,\tau})/V_{i},\forall i\in N^{ES}, \forall t\in T
	\end{equation}
	$Q_{i,t}$ is the cumulative capacity loss of battery $i$ determined by its total throughout $AH_{i,t}$ during the life cycle and the environmental temperature; $V_{i}$ is working voltage of battery $i$. Based on the calculated $Q_{i,t}$, the actual capacity of battery $i$ can be obtained as follows, which is a time-varying parameter. 
	
	\begin{equation}\label{degra-4}
		\overline{E}^{ES,ac}_{i,t} = (1-Q_{i,t})\overline{E}^{ES}_{i},\forall i\in N^{ES}, \forall t\in T
	\end{equation}	
	
	Therefore, the cost of capacity-based battery degradation $C^{ES, de}_{i,n}$ for battery $i$ can be calculated as
	
	\begin{equation}\label{unitcost-SB}
		C^{ES, de}_{i} = N^{re}_{i}p^{ES} + \frac{Q_{i,T^*}}{Q_{i}^{max}}p^{ES}, \forall i \in N^{ES}
	\end{equation}
	where $Q_{i,T^*}$ is the capacity loss of battery $i$ at last time period of BESS life-cycle $T^*$; $Q_{i}^{max}$ is the end-of-life retained capacity percentage (\%); $p^{ES}$ is the unit cost of battery; $N^{re}_{i}$ is the total number of times the battery $i$ needs to be replaced, which is determined as follows.
	
	\begin{equation}
		N^{re}_{i} = \left\{
		\begin{aligned}
			&N^{re}_{i}      &,0 < t < T^*
			\\
			&N^{re}_{i}+1      &,t=T^*
		\end{aligned}
		\right.
	\end{equation}
	
	Considering the replacement of BESS, the degradation cost of BESS during the life-cycle of the grid-connected microgrid is shown in Eq. (\ref{derga-all}).
	
	\begin{equation}
		C^{ES, de} = \sum_{i \in N^{ES}}N^{re}_{i}p^{ES} + \sum_{i \in N^{ES}}\frac{Q_{i,T^*}}{Q_{i}^{max}}p^{ES}
		\label{derga-all}
	\end{equation}

	\subsection{Modeling of Diesel Generators}
	
	According to the role of diesel generators in the microgrid, i.e., fulfilling the power gap between resources and load, the output demand of diesel generators $P^{D}_{t}$ the time period $t$ can be calculated as follows \cite{chen2019optimization}.
	
	\begin{equation}
		P^{D}_{t} = \left\{
		\begin{aligned}
			&0                              &, P_{t}^{Gen}                  \ge     P_{t}^{LD}  \\
			&P_{t}^{LD}-P_{t}^{Gen}         &, P_{t}^{Gen}+N^{DG}P_{i}^{DR}  >       P_{t}^{LD}  \\
			&N^{DG}P_{i,t}^{D}               &, P_{t}^{Gen}+N^{DG}P_{i}^{DR}  \le     P_{t}^{LD}
		\end{aligned}
		\right.
	\end{equation}
	
	\begin{equation}
		P_{t}^{Gen} = \sum_{i \in N^{WT}}P^{W}_{i,t}+\sum_{i \in N^{PV}}P^{P}_{i,t}+\sum_{i\in N^{ES}}(e^{ES}_{i,t}-e^{ES}_{i,t}), \forall t\in T
	\end{equation}
	where $P_{t}^{Gen}$ is the sum of the output power of wind turbines, PV devices and the storage batteries at time period $t$; $N^{DG}$ is the number of diesel generators; $P_{i}^{DR} $ means the rated output power of a diesel generator.
	
	During the operation of the grid-connected microgrid, the power generated from wind and solar has the highest priority to be consumed for promoting renewable energy. Considering the convenience of invoking local resources, when the renewable energy generation equipment cannot meet the demands of the load, the microgrid will give priority to using BESS, and if the BESS still cannot meet the demand, the DG will be activated and followed by the main grid. And if the renewable energy generation exceeds the load demand, the use of BESS is prioritized, followed by the main grid.
	
	\section{Scenario-based Multi-Objective Optimization Model Considering Comprehensive Cost and Carbon Emissions}
	
	As for the grid-connected microgrid, the objective is to minimize comprehensive cost as well as minimize carbon emissions during operation simultaneously via determining DER sizing. The comprehensive cost includes the cost of initial construction, operation, diesel fuel, buying electricity power from the main grid and the profit of selling extra electricity. Among the five types of cost, the last one is special because it is an economic benefit of the grid-connected microgrid. The carbon emissions are quantified by the pollutant emission by $CO_{2}$ (PEC) which represents the environmental objective function.
	
	Combined with the operation characteristics of the microgrid, diminishing carbon emissions means reducing the utilization of DG while increasing the utilization of RES, which will lead to an increase in the comprehensive cost. In this way, a multi-objective optimization approach is employed to weigh different objectives and find out a balanced solution. The result of the multi-objective optimization will converge to a set of solutions that can reflect the trade-offs of objectives and thus provide more diverse choices for microgrid designers to determine the scheme.
	
	Additionally, considering the uncertainty of load demand and power output of RES, i.e., WT and PV, the scenario-based stochastic technique is employed to establish the economic objective and environmental objective. 
	
	\subsection{Economic Objective}
	
	The comprehensive cost of grid-connected microgrid includes the cost of initial construction ($C^{Init}$), the cost of operation and maintenance ($C^{O \& M}$), the cost of diesel fuel due to the operation of DG ($C^{DG}$), the cost of BESS replacement due to degradation effects ($C^{ES,de}$) formulated in (\ref{derga-all}), the cost of buying electricity power from the main grid ($C_{B}^{G}$), and the profit of selling extra electricity to the main grid ($C_{S}^{G}$). Therefore, the comprehensive cost can be represented as in (\ref{SO-E}).
	
	\begin{equation}\label{SO-E}
		f^{E} = C^{Init}+C^{O\&M}+C^{DG}+C^{G,B}-C^{G,S}+C^{ES,de}
	\end{equation}
	where
	
	\begin{equation}\label{SO-E1}
		C^{Init} =N^{WT}p^{WT}+N^{PV}p^{PV}+N^{DG}p^{DG}+N^{ES}p^{ES}
	\end{equation}
	
	\begin{equation}\label{SO-E2}
		\begin{aligned}
			C^{O\&M} &= \sum_{t\in T}N^{WT}p^{WT,O\&M}_{t}+\sum_{t\in T}N^{PV}p^{PV,O\&M}_{t}
			\\
			&+\sum_{t\in T}N^{DG}p^{DG,O\&M}_{t}
		\end{aligned}
	\end{equation}
	
	\begin{equation}\label{SO-E3}
		C^{DG} = \sum_{i\in N^{DG}}\sum_{w \in W}\pi_{w}V^{die}_{i,w}p^{die}
	\end{equation}
	
	\begin{equation}\label{SO-E4}
		C^{G,B} = \sum_{t\in T}\sum_{w\in W}\pi_{w}\lambda_{w,t}^{B}p_{t}^{B}
	\end{equation}
	
	\begin{equation}\label{SO-E5}
		C^{G,S} = \sum_{t\in T}\sum_{w\in W}\pi_{w}\lambda_{w,t}^{S}p_{t}^{S}
	\end{equation}
	
	\begin{equation}\label{SO-E6}
		C^{ES,de} = \sum_{i \in N^{ES}} \sum_{w \in W}\pi_{w}(N^{re}_{i,w}p^{ES} + \frac{Q_{i,T^*,w}}{Q_{i}^{max}}p^{ES})
	\end{equation}
	
	Eq. (\ref{SO-E1})-(\ref{SO-E5}) present the calculation details of five different types of microgrid cost. Among them, $p^{WT}$, $p^{PV}$, $p^{DG}$ and $p^{ES}$ are the unit cost of wind turbines, PV devices, diesel generators and battery storage, respectively. Correspondingly, $N^{WT}$, $N^{PV}$, $N^{DG}$ and $N^{ES}$ are the number of wind turbines, PV devices, diesel generators and batteries, respectively. Parameter $\pi_{w}$ is the probability of occurrence of scenario $w \in W$. Parameter $\lambda_{w}^{B}$ and $\lambda_{w}^{S}$ are the price of buying and selling electricity power with the main grid in the scenario $w$. Correspondingly, $p_{t}^{B}$ and $p_{t}^{S}$ are the value of the amount of electricity traded with the main grid (buying and selling). Parameter $V^{die}_{i,w}$ is the amount of diesel consumed by DG $i$ of scenario $w \in W$ and $p^{die}$ means the price of diesel per liter.
	
	Scenarios describe different load demands as well as power outputs of WT and PV, which would influence the utilization of DG and BESS and the situation of electricity trading with the large power grid. Therefore, the comprehensive cost (\ref{SO-E}) is formed by the weighted sums of $C^{DG}$, $C^{G,B}$, $C^{G,S}$ and $C^{ES}$, in Eq. (\ref{SO-E3})-(\ref{SO-E6}), under all scenarios.
	
	\subsection{Pollutant Emission Objective}
	
	The pollutant emission by $CO_{2}$ (PEC) in the scenario $w$ can be calculated as follows \cite{sheng2017capacity}.
	\begin{equation}\label{ob-pec}
		PEC_{w} = \sum_{t=1}^{T}c^{e}P^{D}_{t,w}
	\end{equation}
	where $T$ is the operation cycle length of microgrid; $c^{e}$ is the coefficient describes the proportional relationship between volume of $CO_{2}$ output power of diesel generators $P^{D}_{t,w}$ in the scenario $w$.
	
	Another environmental objective is given as follows.
	\begin{equation}
		{\rm min} \ f^{P} = \sum_{t\in T}\sum_{w\in W} c^{e}P^{D}_{t,w}
	\end{equation}
	
	Therefore, the objective function of the sizing optimization model is formulated below.
	\begin{equation}
		{\rm{min}} \ F = \{ f^{E}, f^{P} \}
		\label{eq.obfunc}
	\end{equation}
	
	\subsection{Constraints on Microgrid Operation}
	
	Four main constraints need to be considered when we optimize the DER sizing of the grid-connected microgrid under each scenario $w \in W$.
	
	The constraints of the number of DER are shown below:
	\begin{equation}
		\begin{aligned}
			&0<N^{WT}\le N^{WT}_{max}\\
			&0<N^{PV}\le N^{PV}_{max}\\
			&0<N^{DG}\le N^{DG}_{max}\\
			&0<N^{ES}\le N^{ES}_{max}
		\end{aligned}
		\label{const1}
	\end{equation}
	where $N^{WT}_{max}$, $N^{PV}_{max}$, $N^{DG}_{max}$ and $N^{ES}_{max}$ are the maximum number of wind turbines, PV devices, diesel generators and storage batteries respectively \cite{fathy2020recent}.
	
	The constraints of the loss of power supply probability (LPSP) are given below:
	\begin{equation}
		\eta_{L}=\frac{\sum_{t=1}^{T} LPS_{t}}{\sum_{k=1}^{T}P^{LD}_{t}}\le \eta_{Lmax}, \ \forall t
		\label{const2-1}
	\end{equation}
	where $T$ is the service life of the microgrid; $\eta_{Lmax}$ is the maximum allowable LPSP, which is determined by the loss of power supply (LPS) during the service life \cite{fathy2020recent}. The loss of power supply at time period $t$ ($LPS_{t}$) can be calculated as follows:
	\begin{equation}
		LPS_{t} = \left\{
		\begin{aligned}
			&P_{t}^{LD}-P^{Gen,DG}_{t}&, P_{t}^{LD}>P^{Gen,DG}_{t}\\
			&0&, P_{t}^{LD} \le P^{Gen,DG}_{t}
		\end{aligned}
		\right.
		\label{const2-2}
	\end{equation}
	where
	\begin{equation}
		\begin{aligned}
			& P^{Gen,DG}_{t} = \sum_{i \in N^{WT}}P^{W}_{i,t}+\sum_{i \in N^{PV}}P^{P}_{i,t}\\
			& +\sum_{i\in N^{ES}}(e^{ES}_{i,t}-e^{ES}_{i,t})+P_{t}^{D}
		\end{aligned}
	\end{equation}
	
	The operation constraints of the BESS are expressed as follows:
	\begin{equation}
		\begin{aligned}
			&e^{ES}_{i,t} = e^{ES}_{i,t-1}+ \eta^{ch}_{i,t}(N_{i})(e^{ES}_{i,t})p^{ch}_{i,t}-p^{dc}_{i,t}/ \eta^{dc}_{i,t}(N_{i})(e^{ES}_{i,t})
			\\
			&, \forall t\in T, \forall i\in N^{ES}
			\\ 
			&0 \le p^{ch}_{i,t} \le \overline{P^{ch}_{i}},\forall t\in T, \forall i\in N^{ES}
			\\
			&0 \le p^{dc}_{i,t} \le \overline{P^{dc}_{i}},\forall t\in T, \forall i\in N^{ES}
			\\
			&\underline{E}^{ES}_{i} \le e^{ES}_{i,t} \le \overline{E}^{ES,ac}_{i}, \forall t\in T, \forall i\in N^{ES}
			\label{const3}
		\end{aligned}
	\end{equation}
	where constraints (\ref{const3}) impose the limitation on the operation of energy storage units and the charge/discharge power is bounded by their power rating.
	
	Then the operation constraints of DGs are given below:
	\begin{equation}
		\begin{aligned}
			&u_{i,t}\underline{P}^{D}_{i} \le P^{D}_{i,t} \le u_{i,t}\overline{P}^{D}_{i}, \forall i \in N^{DG}, \forall t
			\\
			&P^{D}_{i,t+1}-P^{D}_{i,t} \le u_{i,t}R^{D+}_{i}+(1-u_{i,t})R^{SU}_{i},\forall i \in N^{DG}, \forall t
			\\
			&P^{D}_{i,t-1}-P^{D}_{i,t} \le u_{i,t}R^{D-}_{i}+(1-u_{i,t})R^{SD}_{i},\forall i \in N^{DG}, \forall t
			\\
			&u_{i,t},v^{SU}_{i,t},v^{SD}_{i,t}\in\{0,1\},\forall i \in N^{DG}, \forall t
			\\
			&v^{SU}_{i,t}+v^{SD}_{i,t}\le1,\forall i \in N^{DG}, \forall t
			\\
			&u_{i,t+1} = u_{i,t}+v^{SU}_{i,t}-v^{SD}_{i,t},\forall i \in N^{DG}, \forall t  
		\end{aligned}
		\label{const4}
	\end{equation}
	where the binary variables $u_{i,t}$, $v_{i,t}^{SU}$, $v_{i,t}^{SD}$ represent the on/off status, start-up action and shut-down action of DG $i$ at time period $t$.
	
	System power balance is enforced as follows:
	\begin{equation}
		\begin{aligned}
			&\sum_{i \in N^{WT}} P^{W}_{i,t}+\sum_{i \in N^{PV}} P^{P}_{i,t}+\sum_{i \in N^{DG}} P^{D}_{i,t}+\sum_{i \in N^{ES}} p^{dc}_{i,t}
			\\ 
			&= \sum_{i \in N^{ES}} p^{ch}_{i,t}+P^{G}_{t}+P^{LD}_{t}, \forall t
		\end{aligned}
		\label{const5}
	\end{equation}
	
	Overall, the entire multi-objective optimization model can be established as follows:
	\begin{equation}\label{model}
		\begin{aligned}
			{\min} \quad & F = \{ f^{E}, f^{P} \}
			\\
			\text{s.t.} \quad
			&(\ref{SO-E1}), (\ref{SO-E2}), (\ref{SO-E3}), (\ref{SO-E4}), (\ref{SO-E5}), (\ref{SO-E6}), (\ref{ob-pec})
			\\  
			&(\ref{degra-2}), (\ref{degra-3}), (\ref{degra-4}), (\ref{derga-all})
			\\
			&(\ref{const1}), (\ref{const2-1}), (\ref{const2-2}), (\ref{const3}), (\ref{const4}), (\ref{const5})
		\end{aligned}
	\end{equation}

	\section{Self-Adaptive Multi-Objective Genetic Algorithm}
	Because of the nonlinear degradation effects of BESS described in Eq. (\ref{degra-2})-(\ref{degra-4}), the stochastic multi-objective sizing optimization (SMOSO) model of grid-connected microgrid (\ref{model}) is the Mixed Integer Nonlinear Program (MINLP), which cannot be directly solved by commercial solvers such as CPLEX or Gurobi. This paper proposed a new self-adaptive multi-objective genetic algorithm (SAMOGA) to solve the SMOSO model and determine the optimal sizing of DER in the microgrid.
	
	The SAMOGA increases the performance of searching optimal DER sizing configurations from two aspects, i.e., the pre-grouped hierarchical selection and the self-adaptive probabilities of cross and mutation, which improves the local optimal defect of the traditional genetic algorithm. The two main stages, i.e., selection and crossover \& mutation are described in the following sections.
	
	\begin{figure}[!t]
		\renewcommand{\algorithmicrequire}{\textbf{Input:}}
		\renewcommand{\algorithmicensure}{\textbf{Output:}}
		\begin{algorithm}[H]
			\caption{Pre-Grouped Hierarchical Selection}
			\begin{algorithmic}[1]
				\REQUIRE The groups of individuals $N_{G}$, the population size $N$, the set of individuals fitness $Fit(S)$.        
				\ENSURE  The selected individuals $j_{1}$, $j_{2}$.  
				
				\STATE {Initialize the number of selected individuals, i.e. $j = 1$.}
				\STATE {$S^{*}$ $\leftarrow$ Sort ($S$) based on $Fit(S)$.}
				\STATE {$N_{G}$ $\leftarrow$ Divide ($S$) based on $S^{*}$.}
				\STATE {$Fit(G)$ $\leftarrow$ Fitness ($N_{G}$).}
				\WHILE{$j \le 2$}{
					\STATE {$Group$  $\leftarrow$ Select ($N_{G}$) based on $Fit(G)$.}
					\STATE {$Individual$ $\leftarrow$ Select ($Group$) randomly.}
					\STATE {$j$ $\leftarrow$ $j+1$}}
				\ENDWHILE
			\end{algorithmic}
			\label{alg.2}
		\end{algorithm}
	\end{figure}
	
	\begin{figure}[!t]
		\renewcommand{\algorithmicrequire}{\textbf{Input:}}
		\renewcommand{\algorithmicensure}{\textbf{Output:}}
		\begin{algorithm}[H]
			\caption{Self-Adaptive Multi-Objective Genetic Algorithm (SAMOGA)}
			\begin{algorithmic}[1]
				\REQUIRE Maximum number of iterations $i_{m}$, the population size $N$, initial probabilities of cross $P_{c,0}$, initial probabilities of mutation  $P_{m,0}$.        
				\ENSURE The optimal sizing of DER, i.e., WT, PV, DG and BESS.  
				
				\STATE {The set of individuals $S$ $\leftarrow$ Encode ($Population$).}
				\STATE {The number of iteration $i$ $\leftarrow$ 1.}
				\WHILE {$i\le i_{m}$}{
					\STATE {$Fit(S)$ $\leftarrow$ Fitness ($S$).}
					\STATE {Update $P_{c}$ and $P_{m}$ via Eq.~(\ref{aeq.1}) and Eq.~(\ref{aeq.2})}
					\STATE {$n$ $\leftarrow$ 0, $S$ $\leftarrow$ $\emptyset$}
					\WHILE {$ n \le N$}{
						\STATE {$j_{1}$, $j_{2}$ $\leftarrow$ HierarchicalSelection ($N$,$N_{G}$,$Fit(S)$).}
						\STATE {$Subpop$ $\leftarrow$ $j_{1}$ $\bigcup$ $j_{2}$}
						\STATE {$Subpop$ $\leftarrow$ Crossover ($Subpop$) via $P_{c}$.}
						\STATE {$Subpop$ $\leftarrow$ Mutation ($Subpop$)  via $P_{m}$.}
						\STATE {$S$ $\leftarrow$ $S$ $\bigcup$ $Subpop$, $n$ $\leftarrow$ $n+2$}
					}
					\ENDWHILE
					\STATE {$i$ $\leftarrow$ $i+1$.}
				}
				\ENDWHILE
				\STATE {Choose the individuals which are on the Pareto optimal front in the last generation as the optimal results calculated by SAMOGA.}
				
			\end{algorithmic}
			\label{alg.1}
		\end{algorithm}
	\end{figure}

	\subsection{Pre-Grouped Hierarchical Selection}
	
	The dominance of the individuals who have higher fitness reflects on the higher probability of being selected during the evolutionary period of the heuristic genetic algorithm, which will lead to premature local optimal solutions. In this section, the pre-grouped hierarchical selection method is deployed to decrease the dominance of the individuals who have higher fitness via adding appropriate extra randomness during the selection stage \cite{falcon2020indicator}. 
	
	The main steps of pre-grouped hierarchical selection are presented in Algorithm \ref{alg.2}. The operators Fitness ($\cdot$), Sort ($\cdot$), Divide ($\cdot$), and Select ($\cdot$) are defined to calculate the fitness of population or groups via objective functions, sort population, divide the population into groups and select groups or individuals via roulette method.
	
	\subsection{Self-Adaptive Probabilities of Crossover and Mutation}
	
	\begin{figure}
		\centering
		\includegraphics[width=0.95\linewidth]{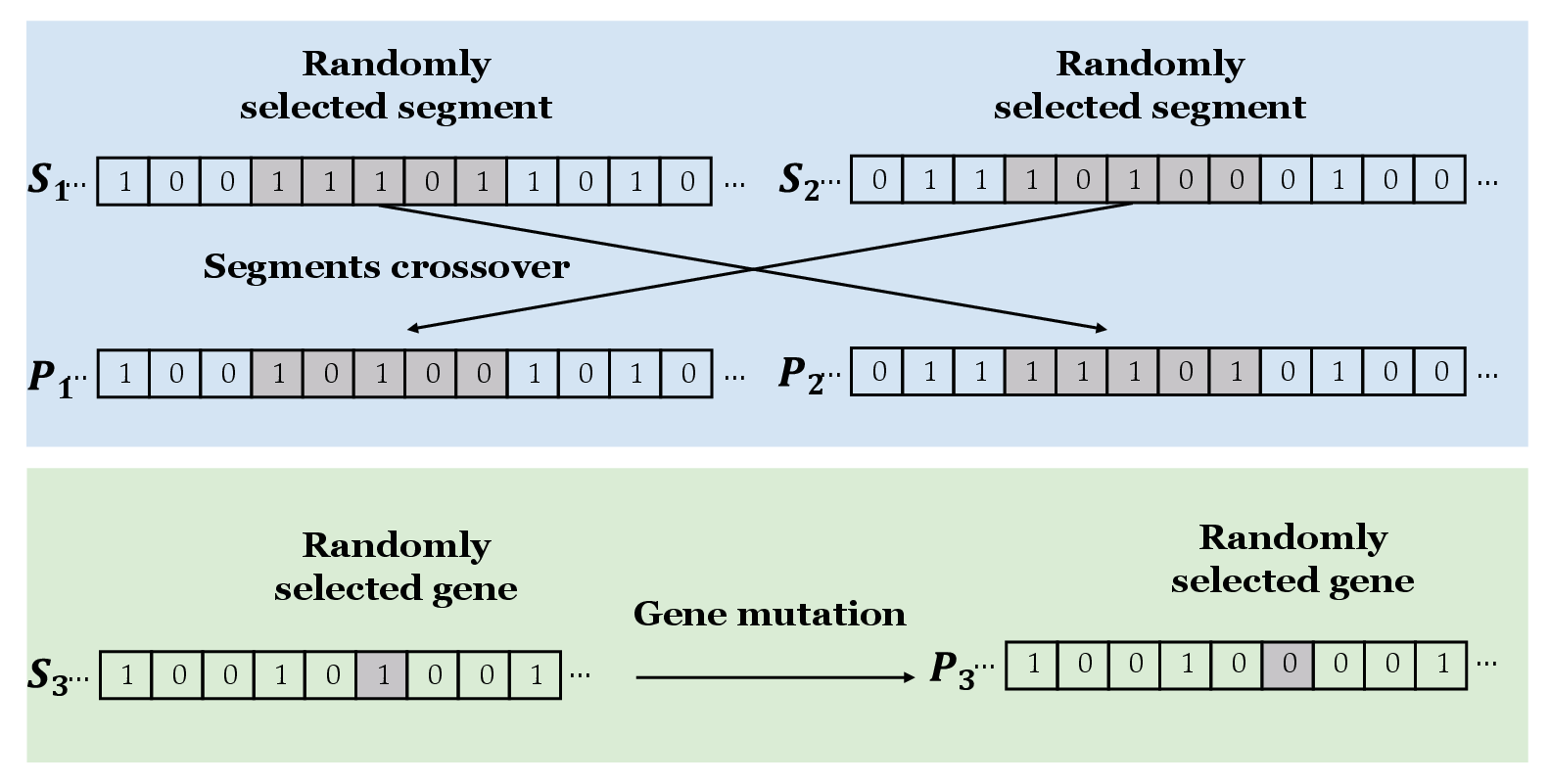}
		\caption{Illustration of the crossover and mutation operation.}
		\label{fig.cross}
	\end{figure}
	
	Apart from the selection operation, crossover and mutation are another two operations of GA that are of importance to the searching performance as illustrated in Fig. \ref{fig.cross}. During the early stage of evolution, the population needs a faster speed of generating new individuals as well as avoiding random searching, which needs a higher probability of crossover and a lower probability of mutation, respectively. During the middle and late stages of evolution, the population aims to maintain the current optimum of the population as well as generate elite individuals close to the global optimum, which needs a lower probability of crossover and a higher probability of mutation, respectively.
	
	Based on the above requirements, the self-adaptive probabilities of cross $P_{c}$ and mutation $P_{m}$ are employed. Compared with the adaptive strategy in \cite{AGA-1}, Eq.(\ref{aeq.1}) - Eq.(\ref{aeq.2}) fully consider the degree of convergence ($g_{c}$) of the whole population instead of just individuals to enhance the dynamic adaptability of crossover and mutation \cite{ooi2019self}.
	
	\begin{equation}
		P_{c} = \frac{P_{c,0}}{1+\alpha\frac{{\rm lg}[{(g+g_{c})}]} {i_{m}}}, g = 1,2,...,i_{m}
		\label{aeq.1}
	\end{equation}
	
	\begin{equation}
		P_{m} = P_{m,0} (1+\beta\frac{{\rm lg}(g+g_{c})}{i_{m}}), g = 1,2,...,i_{m}
		\label{aeq.2}
	\end{equation}
	where $P_{c,0}$ and $P_{m,0}$ are the initial probabilities of cross and mutation respectively; $g$ is the current number of iterations; $g_{c}$ means the number of iterations where optimal fitness remain unchanged; $i_{m}$ is the maximum number of iterations; $\alpha$ and $\beta$ are the regulated parameters.
	
	Compared with traditional genetic algorithms, i.e. HSGA-II \cite{NSGA} and AGA \cite{AGA}, with the crossover and mutation probabilities of constant or linear change, the proposed SAMOGA can adaptively change the probabilities according to the degree of population convergence. In this way, the evolution is faster and more directionally in the early stage, and the current optimum of the population can be maintained more effectively in the middle or late stage. The complete process of SAMOGA is summarized in Algorithm \ref{alg.1} where the HierarchicalSelection ($\cdot$) represents the selection process of Algorithm \ref{alg.2}.

	\section{Case Study}
	
	\subsection{Setup}
	
	\begin{figure*}
		\centering
		\subfloat[]{
			\includegraphics[width=0.33\linewidth]{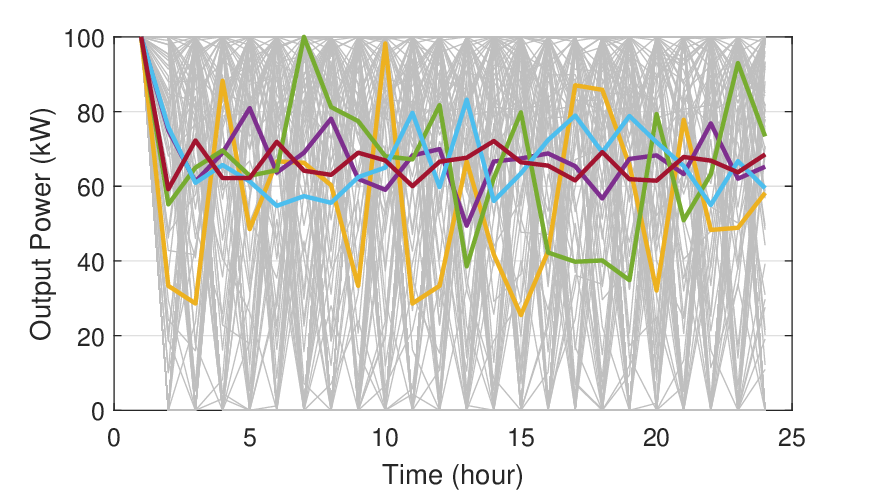}}
		\subfloat[]{
			\includegraphics[width=0.33\linewidth]{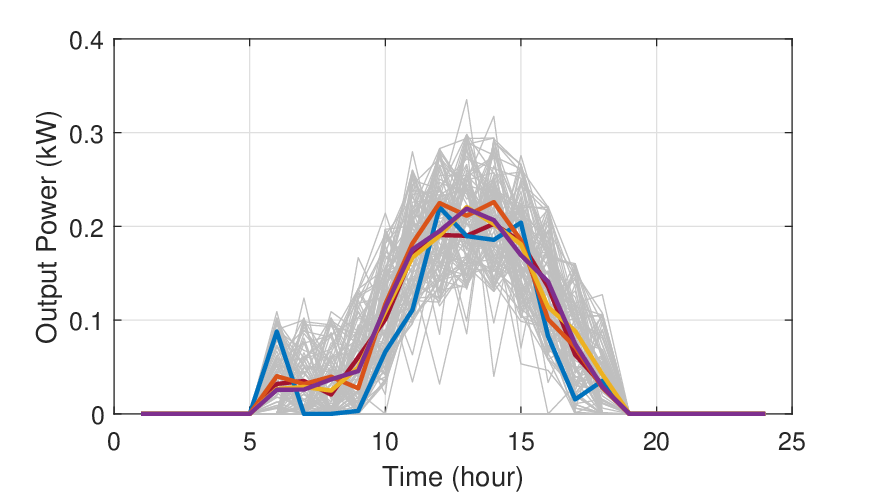}}
		\subfloat[]{
			\includegraphics[width=0.33\linewidth]{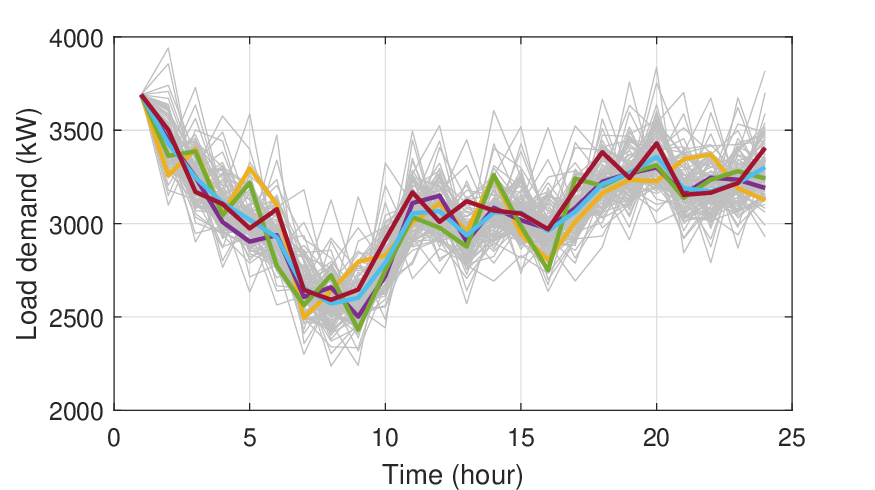}}
		\caption{Each scenario contains generations of WT and PV as well as the load demand and the total scenario number is 125 ($5^{3}$): (a) 5 typical scenarios of WT's output power (kW); (b) 5 typical scenarios of PV's output power (kW); (c) 5 typical scenarios of load requirement (kW).}
		\label{fig.scenario}
	\end{figure*}
	
	In the case studies, the stochastic multi-objective sizing optimization model (\ref{model}) is solved by the proposed SAMOGA so as to obtain the optimal sizing configurations of the microgrid. The scenario data including generation of WT, PV, and load demand are generated through Latin hypercube sampling and reduced through K-means clustering algorithm \cite{LHS,Kmeans} as shown in Fig. \ref{fig.scenario}. The initial data about the generations of WT and PV are calculated through the model (\ref{eq.1}) and (\ref{eq.2}) based on the operation parameters in \cite{parameter-new1, fathy2020recent}. A total of 125 stochastic scenarios are generated and randomly select 10, 20, and 30 scenarios to analyze. 
	
	\begin{table}[htbp]
		\centering
		\caption{Operation and economic parameters of DERs}
		\setlength{\tabcolsep}{5mm}
		\begin{tabular}{rcc}
			\toprule
			\multicolumn{1}{c}{Component} & Parameter & Value \\
			\midrule
			\multicolumn{1}{c}{WT} 
			& $v_{cin}$  & 3 $m/s$ \\
			& $v_{cout}$ & 25 $m/s$ \\
			& $v_{r}$    & 12 $m/s$ \\
			& $P^{WR}$   & 100 $kW$  \\
			& $p^{WT}$    & 100000 \$  \\
			& $p^{WT,O\&M}_{t}$   &  1.14 \$/$h$  \\
			\midrule 
			\multicolumn{1}{c}{PV} 
			& $G_{stc}$  & 1 $kW/m^{2}$ \\
			& $k_{p}$    & -0.40 \% \\
			& $T_{en}$  & 290 $K$ \\
			& $P^{PR}$   & 0.33 $kW$ \\
			& $p^{PV}$    & 400 \$ \\
			& $p^{PV,O\&M}_{t}$   &  0.0057 \$/$h$  \\
			\midrule
			\multicolumn{1}{c}{DG}
			& $P^{DR}$   & 500 $kW$ \\
			& $p^{DG}$    & 40000 \$ \\
			& $p^{DG,O\&M}_{t}$   & 0.0685 \$/$h$ \\
			& $c_{n}$    & 232.04 $g/kWh$ \\
			&  $p_{diesel}$ & 1.11 \$/$L$ \\
			\midrule
			\multicolumn{1}{c}{BESS} 
			& $\eta_{init}$  & 96.1 \% \\
			& $\underline{E}^{ES}_{i}$  & 5 $kWh$ \\
			& $\overline{E}^{ES}_{i}$   & 50 $kWh$    \\
			& $p^{ES}$    & 10000 \$ \\
			\bottomrule
		\end{tabular}%
		\label{tab:para}%
	\end{table}%
	
	The pre-grouped number $N_{G}$, the maximum iterative number $i_{m}$, the population size $N$, initial probabilities of cross $P_{c,0}$ and mutation $P_{m,0}$, parameters $\alpha$ and $\beta$ are set as 5, 50, 30, 0.65, 0.01, 10 and 10 respectively. The maximum number of WT, PV, DG, batteries and allowable LPSP are set as 31, 16383, 15, 255 and 40\%. The parameter values of  $\kappa$, $E_{a}$, $R$, $z$ and $V$ are set as 19300, -31000, 8.314, 0.554 and 240, respectively \cite{degradation-3}.
	
	The operation and economic parameters of DERs are described in TABLE \ref{tab:para} which contains four main components of the microgrid i.e. WT, PV, DG and BESS. In this section, case studies are conducted on MATLAB with a laptop with Intel Core i7-10700 2.90GHz CPU and 16GB of RAM. 
	
	\subsection{Life Cycle Sizing Optimization of Microgrid}
	
	\begin{table}[htbp]
		\centering
		\caption{Optimized Sizing of Microgrid over One-year Life Cycle}
		\setlength{\tabcolsep}{0.015\columnwidth}
		\begin{tabular}{ccccccc}
			\toprule
			Solution & \multirow{2}[4]{*}{Cost/\$} & \multirow{2}[4]{*}{PEC/kg} & \multicolumn{4}{c}{Optimized configuration} \\
			\cmidrule{4-7}     number &       &       & WT    & DG    & BESS  & PV \\
			\midrule
			1     & 7858551 & 3276596 & 31    & 8     & 2     & 748 \\
			2     & 7859055 & 3276251 & 31    & 8     & 2     & 750 \\
			3     & 8753034 & 2736055 & 31    & 8     & 2     & 4110 \\
			4     & 8963295 & 2631548 & 31    & 8     & 2     & 4844 \\
			5     & 9484528 & 2408236 & 31    & 8     & 2     & 6572 \\
			6     & 10265220 & 2169157 & 31    & 8     & 2     & 8932 \\
			7     & 11177141 & 1996843 & 31    & 8     & 16    & 10670 \\
			8     & 11449801 & 1955165 & 31    & 8     & 0     & 12302 \\
			9     & 11494913 & 1946760 & 31    & 8     & 2     & 12292 \\
			10    & 11498726 & 1946242 & 31    & 8     & 2     & 12302 \\
			11    & 11507160 & 1945086 & 31    & 8     & 2     & 12324 \\
			12    & 11799539 & 1900275 & 31    & 8     & 16    & 12302 \\
			13    & 11808048 & 1899107 & 31    & 8     & 16    & 12324 \\
			14    & 11924214 & 1883643 & 31    & 8     & 22    & 12334 \\
			15    & 12083939 & 1863485 & 31    & 8     & 16    & 13038 \\
			16    & 15140305 & 1575922 & 31    & 8     & 220   & 12302 \\
			\bottomrule
		\end{tabular}%
		\label{tab:result}%
	\end{table}%

	\begin{figure}
		\centering
		\includegraphics[width=0.95\linewidth]{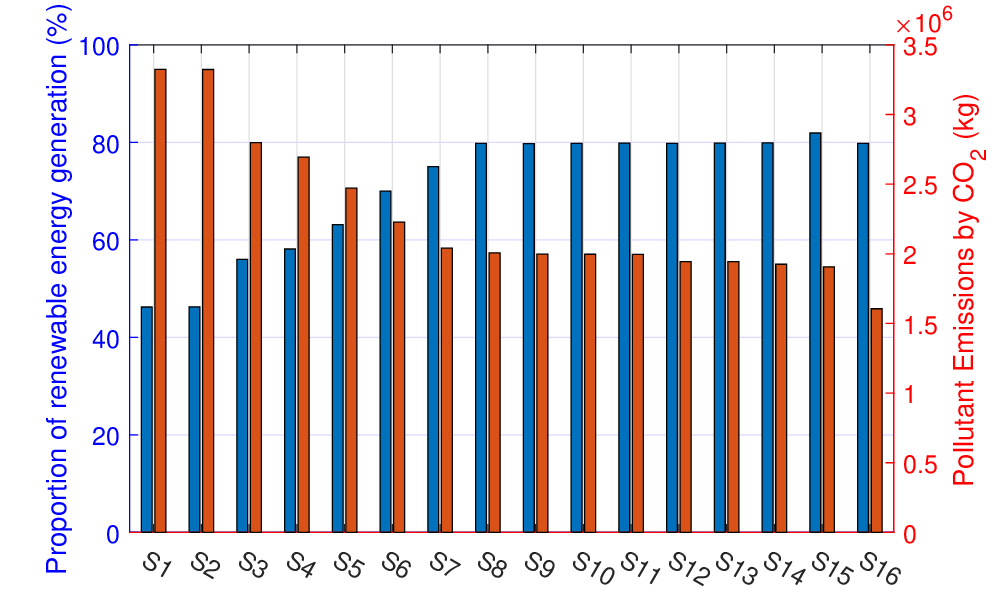}
		\caption{The histogram describes the proportion of renewable energy generation (\%) in blue with the left axis and PEC ($kg$) in red with the right axis of different solutions in the Pareto optimal frontier (S1 is the abbreviation of solution 1).}
		\label{fig.bar}
	\end{figure}

	TABLE \ref{tab:result} shows the optimized sizing configurations of the microgrid and the corresponding values of comprehensive cost and pollutant emissions by $CO_{2}$ (PEC). Among the sixteen configurations of three types of generators and BESS, it can be seen that after considering the carbon emissions in the sizing optimization of microgrid, the value of PEC has shown a marked decline while the numbers of PV devices are shown a considerable increasing trend from configuration 1 to 16. It means the additional volume of RES actually reduces carbon emissions. This is also reflected in the proportion of renewable energy generation calculated according to Eq.~(\ref{eq.Pr}) which increases gradually with the decline of the value of PEC in Fig. \ref{fig.bar}. 	Specifically, from configuration solution-8 to solution-16, the proportion of renewable energy generation is more than 79\%, which means RES occupies a dominant position on the power generation side of the microgrid. 
	
	\begin{equation}
		Pr_{RES} = \frac{\sum_{t\in T}\sum_{i\in N^{WT}}P_{i,t}^{W}+\sum_{t\in T}\sum_{i\in N^{PV}}P_{i,t}^{P}}{\sum_{t\in T}P_{t}^{LD}}
		\label{eq.Pr}
	\end{equation}
	
	\begin{figure}
		\centering
		\includegraphics[width=1\linewidth]{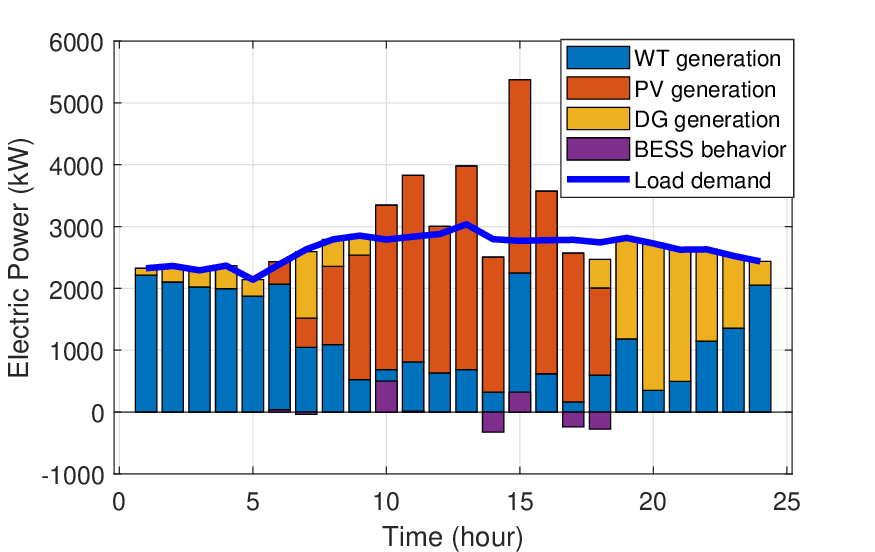}
		\caption{The operations of DERs in one day including generation of WT, PV and DG as well as BESS behavior (charging if positive and discharging if negative).}
		\label{fig.typicalday}
	\end{figure}
	
	\begin{figure}
		\centering
		\includegraphics[width=0.95\linewidth]{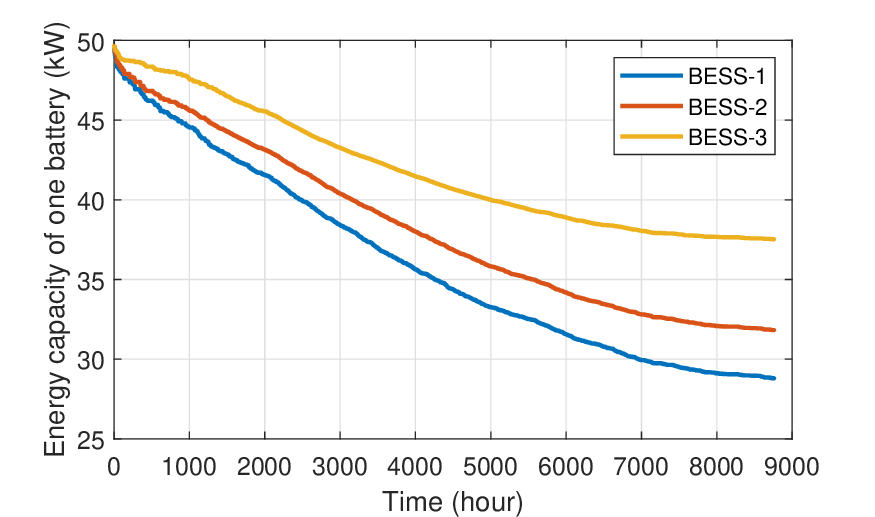}
		\caption{The analysis of BESS capacity over one-year cycle: BESS-1 corresponds to Solution-10, BESS-2 corresponds to Solution-12 and BESS-3 corresponds to Solution-16.}
		\label{fig.BESS}
	\end{figure}

	Take the microgrid scheduling strategy of one day in the year to illustrate the relationship between DER and load demand. In Fig. \ref{fig.typicalday}, during the time period [6,10,11,15], the sum of PV generation and WT generation is more than load demand, therefore BESS is activated in charging mode. While during the time period [7,14,17,18], BESS is activated in discharging mode to maintain the stability of the microgrid. 
	
	Fig. \ref{fig.BESS} indicates the change of BESS capacity during operation, it can be seen that BESS-1 (solution-10), BESS-2 (solution-12) and BESS-3 (solution-16) all present the nonlinear declines in capacity. However, BESS-3 shows less capacity retention because larger BESS sizing can reduce the degradation of battery cells under the same conditions. Ignoring the degradation of BESS will cause inaccurate evaluation of the microgrid, i.e., 1.71\% error on cost and 0.43\% error on PEC averagely. And the calculation error may continue to magnify with the increase of the life cycle and expansion of the scale of the microgrid.
	
	\begin{figure}
		\centering
		\includegraphics[width=0.95\linewidth]{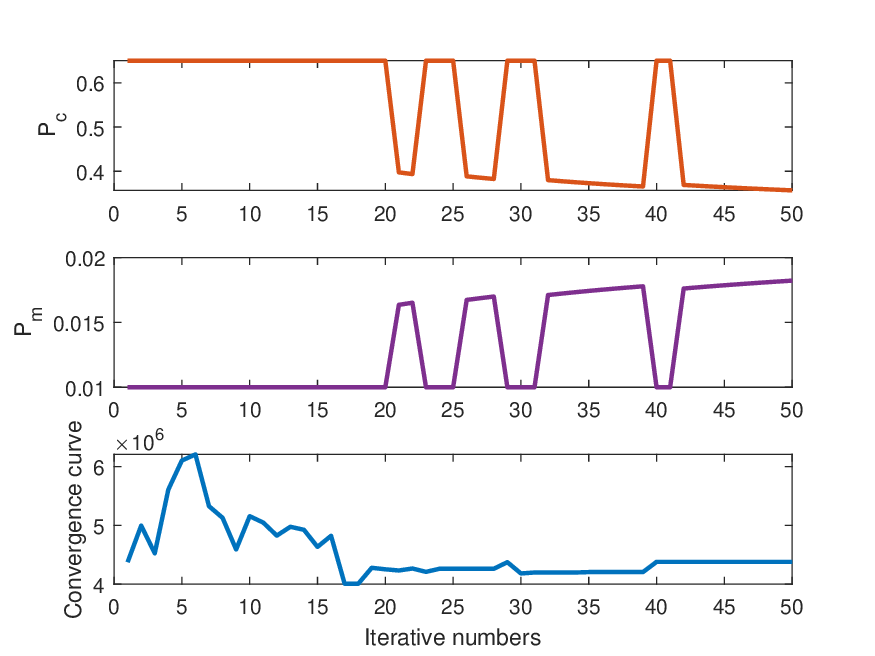}
		\caption{The convergence curve of SAMOGA and corresponding probabilities of crossover and mutation.}
		\label{fig.iter}
	\end{figure}

	Fig. \ref{fig.iter} shows the convergence curve of SAMOGA and corresponding probabilities of crossover and mutation. It can be seen that after the iteration number reaches 20, the evolution enters local optimality. However, the probabilities of crossover and mutation change adaptively with the convergence degree of the population. The adaptive probabilities help the population jump out of the local optimality when iteration number = 23, 29 and 40, which enhances the optimum searching performance and population diversity.

	\begin{figure*}
		\centering
		\subfloat[]{
			\includegraphics[width=0.33\linewidth]{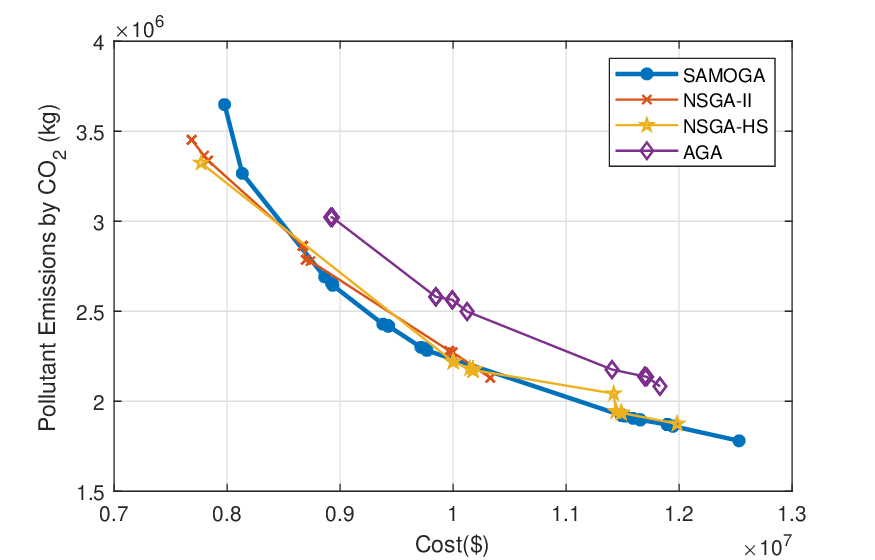}}
		\subfloat[]{
			\includegraphics[width=0.33\linewidth]{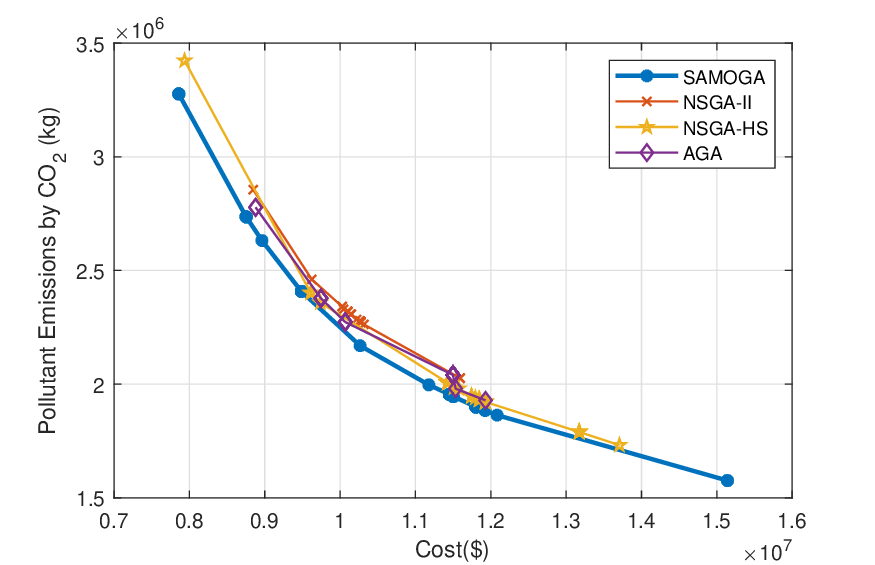}}
		\subfloat[]{
			\includegraphics[width=0.33\linewidth]{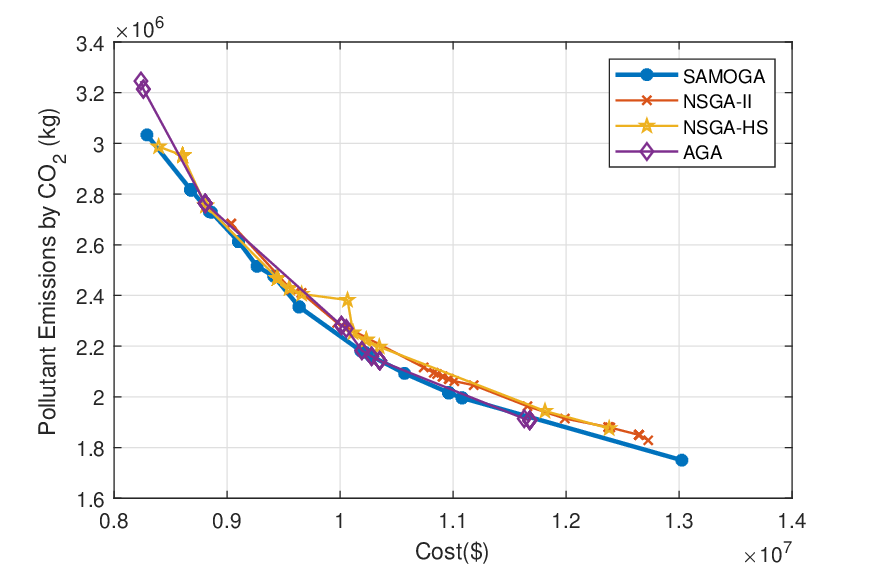}}
		\caption{Pareto frontiers of NSGA-II, HSGA-HS, AGA and SAMOGA under different numbers of stochastic scenarios: (a) 10 scenarios; (b) 20 scenarios; (c) 30 scenarios.}
		\label{fig.Pareto-scenario}
	\end{figure*}

	\subsection{Comparative Performance Study}
	
	In order to illustrate the effectiveness of SAMOGA proposed in this paper, we compare the Pareto optimal frontiers of SAMOGA and other effective genetic algorithms in Fig. \ref{fig.Pareto-scenario} under the same conditions. The summaries of the comparative algorithms are:
	\begin{enumerate}
		\item NSGA-II \cite{NSGA}: A fast and elitist multi-objective genetic algorithm with constant crossover and mutation probabilities and tournament selection method.
		\item NSGA-HS \cite{NSGA}: Modified NSGA-II with the pre-grouped hierarchical selection method.
		\item AGA \cite{AGA}: Adaptive genetic algorithm with crossover and mutation probabilities of linear change.
	\end{enumerate}
	
	NSGA-HS is the modified NSGA-II algorithm with the pre-grouped hierarchical selection method which can decrease the dominance of the individuals who have higher fitness via adding appropriate extra randomness during the selection stage, their performance comparison will be discussed in this section. It can be seen in Fig. \ref{fig.Pareto-scenario} that the Pareto frontiers of SAMOGA are generally located in the lower left region of the coordinate axis compared with the other three methods under different scenarios.

	\subsubsection{Optimum Comparative}
	
	\begin{table}[htbp]
		\centering
		\caption{Largest ORAs of four algorithms under various scenario numbers ($\times 10^{13}$)}
		\begin{tabular}{ccccc}
			\toprule
			Scenario numbers & NSGA-II & NSGA-HS & AGA   & SAMOGA \\
			\midrule
			10    & 1.061  & 1.068  & 0.882  & \textbf{1.071 } \\
			20    & 0.995  & 1.028  & 1.024  & \textbf{1.050 } \\
			30    & 1.029  & 1.028  & 1.055  & \textbf{1.058 } \\
			\bottomrule
		\end{tabular}%
		\label{tab:compare1}%
	\end{table}%
	
	\begin{figure}
		\centering
		\includegraphics[width=0.8\linewidth]{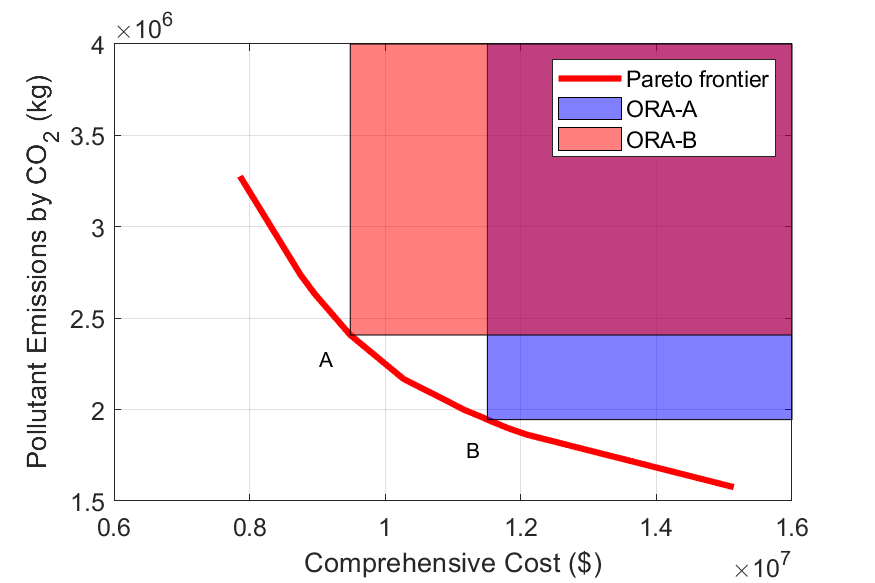}
		\caption{The Pareto frontier and its objective rectangle areas (ORAs).}
		\label{fig.ORA}
	\end{figure}

	According to the theory of Nash bargaining solution \cite{Nashbargain}, the larger the objective rectangle areas (ORAs), the farther the corresponding solution on the Pareto frontier is from the worst cases. That means the comprehensive optimum of various solutions in the Pareto frontier can be quantified as in Eq. (\ref{eq.nash}). 
	\begin{equation}
		F = (C^*-f^{E})\cdot(P^*-f^{P})
		\label{eq.nash}
	\end{equation}
	where $C^*$ and $P^*$ are the worst cases of cost and PEC respectively.
	
	As in Fig. \ref{fig.ORA}, if ORA-A is larger than ORA-B, then the comprehensive optimum of solution-A is better than that of solution-B. Therefore, we compare the largest ORAs of four methods under different scenario numbers in TABLE \ref{tab:compare1} to verify the performance of searching optimum individuals.
	
	The worst cases used to calculate the ORAs are set as $1.6 \times 10^{7}(\$)$ for cost and $4 \times 10^{6}(kg)$ for PEC, respectively. TABLE \ref{tab:compare1} shows that SAMOGA is better than NSGA-II, NSGA-HS and AGA in best individual searching performance. SAMOGA improves three algorithms by 7.55\%, 3.40\% and 2.01\% under 10, 20 and 30 scenarios averagely, which verifies the effectiveness and superiority of SAMOGA.

	\begin{table}[htbp]
		\centering
		\caption{Number of diverse solutions of four algorithms on cost and PEC under various scenario numbers}
		\begin{tabular}{cccccc}
			\toprule
			Scenario &       & \multirow{2}[2]{*}{NSGA-II} & \multirow{2}[2]{*}{NSGA-HS} & \multirow{2}[2]{*}{AGA} & \multirow{2}[2]{*}{SAMOGA} \\
			numbers &       &       &       &       &  \\
			\midrule
			\multirow{2}[2]{*}{10 } & Cost  & 5     & 5     & 7     & \textbf{9 } \\
			& PEC   & 7     & 6     & 6     & \textbf{10 } \\
			\midrule
			\multirow{2}[1]{*}{20 } & Cost  & 6     & 9     & 5     & \textbf{11 } \\
			& PEC   & 7     & 9     & 6     & \textbf{10 } \\
			\midrule
			\multirow{2}[1]{*}{30 } & Cost  & 11    & 11    & 6     & \textbf{12 } \\
			& PEC   & 12    & 12    & 7     & \textbf{13 } \\
			\bottomrule
		\end{tabular}%
		\label{tab:compare2}%
	\end{table}%
	
	\subsubsection{Diversity Comparative}
	
	Apart from the optimum of individuals, the diversity of the population is another important index of genetic algorithm \cite{GA-variance}. We introduce the definition of ``number of diverse solutions'' that represents the total number of solutions on the Pareto frontier whose cost or PEC differences between each other are greater than certain values, i.e., $10^{5}~(\$)$ on cost and $2\times 10^{4}~(kg)$ on PEC respectively. Therefore, we can compare the solution diversity of four algorithms on the Pareto frontier quantitatively based on the number of diverse solutions. 
	
	As shown in TABLE \ref{tab:compare2}, the number of diverse solutions of each algorithm is calculated. It can be seen that SAMOGA can obtain more diverse solutions both in cost and PEC dimension compared with other algorithms under scenario numbers 10, 20 and 30, which means that SAMOGA can provide more diverse choices for microgrid sizing optimization. Additional comparisons between NSGA-II and NSGA-HS on optimum and diversity also validate the advancement of the pre-grouped hierarchical selection approach.
	
	\subsection{Sensitivity Analysis}
	\begin{figure}
		\centering
		\subfloat[]{
			\includegraphics[width=0.5\linewidth]{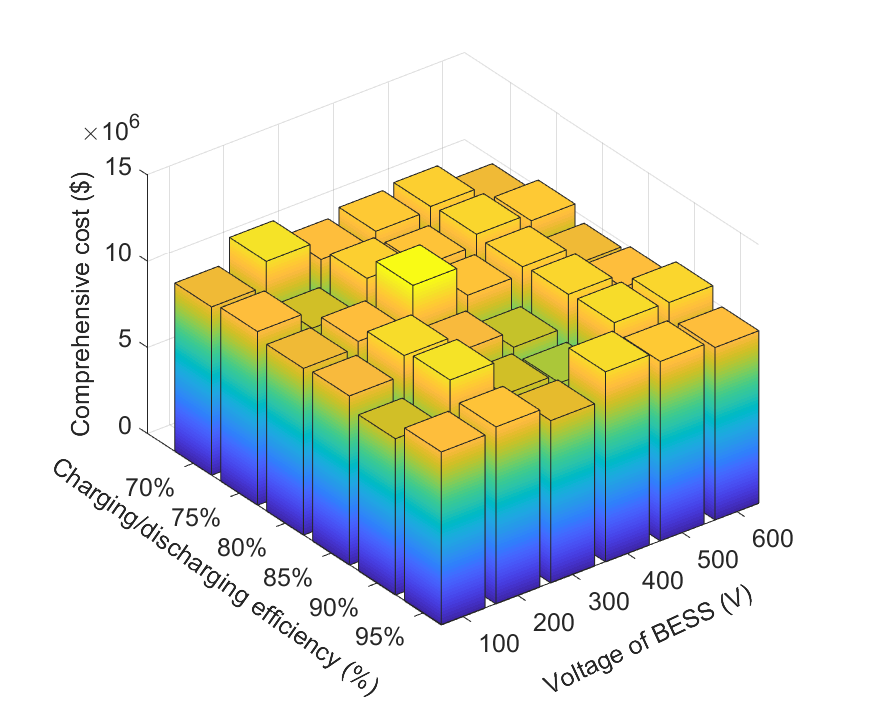}}
		\subfloat[]{
			\includegraphics[width=0.5\linewidth]{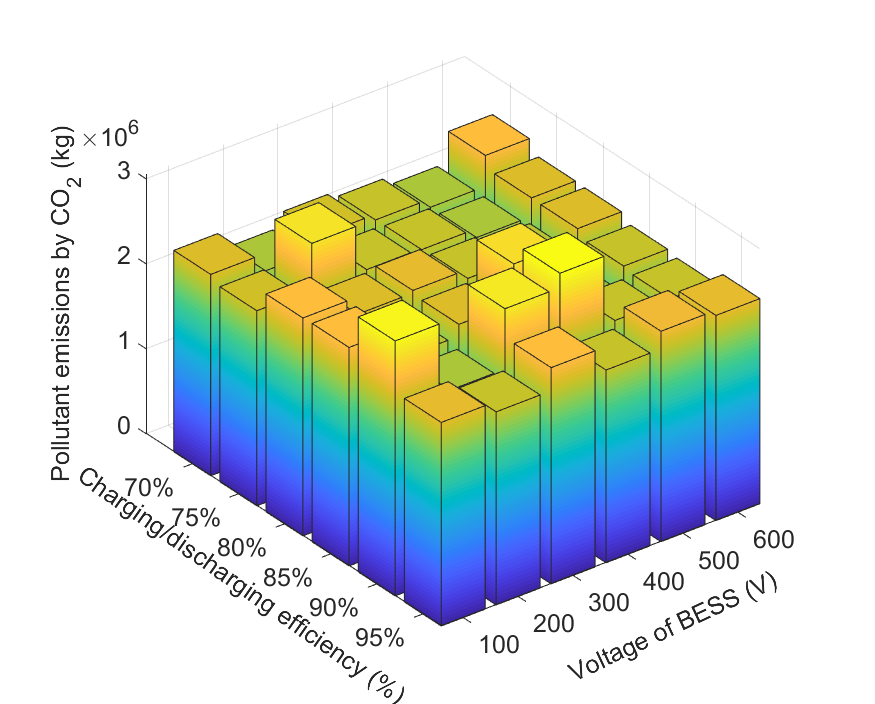}}
		\caption{The impact of charging/discharging efficiency and operation voltage of BESS on comprehensive cost (a) and pollutant emissions by $CO_{2}$ (b).}
		\label{fig.Pareto-sen}
	\end{figure}
	
	In this section, the impact of charging/discharging efficiency taking values from 70\% to 95\% with a gradient of 5\%, and that of operation voltage of BESS taking values from 100V to 600V with a gradient of 100V are studied (Fig. \ref{fig.Pareto-sen}). With the change of parameters, the robustness of the comprehensive cost is higher slightly compared with that of pollutant emissions by $CO_{2}$ (PEC). In detail, with the change of efficiency in 5\%, the amplitude of change of comprehensive cost is 7.2\% while 8.53\% of PEC; with the change of voltage in 100V, the amplitude of change of comprehensive cost is 10.28\% while 11.02\% of PEC.

	\section{Conclusion}
	
	This paper proposes a scenario-based multi-objective optimization model for grid-connected microgrid considering the cost and carbon emissions to realize the optimization of economy-environmental comprehensive benefits. The nonlinear degradation effect of BESS is considered and a new self-adaptive multi-objective genetic algorithm (SAMOGA) is proposed to solve the nonlinear optimization model. Case studies suggest that:
	
	\begin{enumerate}
		\item There is a significant trade-off relationship between comprehensive cost and pollutant emissions by $CO_{2}$ (PEC). The proportion of more than 79\% renewable energy generation can effectively keep PEC at a low level.
		\item The degradation effect of BESS is related to the sizing of batteries configured and larger BESS sizing can reduce the degradation under the same conditions. Besides, ignoring the degradation of BESS will cause an inaccurate evaluation of the microgrid, i.e., 1.71\% error on cost and 0.43\% error on PEC average.
		\item The proposed SAMOGA has better performance on both the optimum and diversity of the solutions. SAMOGA improves three comparative algorithms by 7.55\%, 3.40\%, and 2.01\% under 10, 20, and 30 scenarios average on the optimum comparison. On diversity comparison, SAMOGA can obtain more diverse solutions both in cost and PEC dimension. 		
	\end{enumerate}
	
	In our future work, microgrid planning will be considered in the context of real-world electricity markets. Besides, the uncertainty characterization of renewable energy will be conducted based on data-driven approaches.

	\section*{Acknowledgements}
	
	The authors would like to thank the editor and anonymous reviewers for their insightful comments and suggestions.

	\footnotesize
	\bibliography{reference}
	\bibliographystyle{IEEEtran}

	\begin{IEEEbiography}[{\includegraphics[width=1in,clip,keepaspectratio]{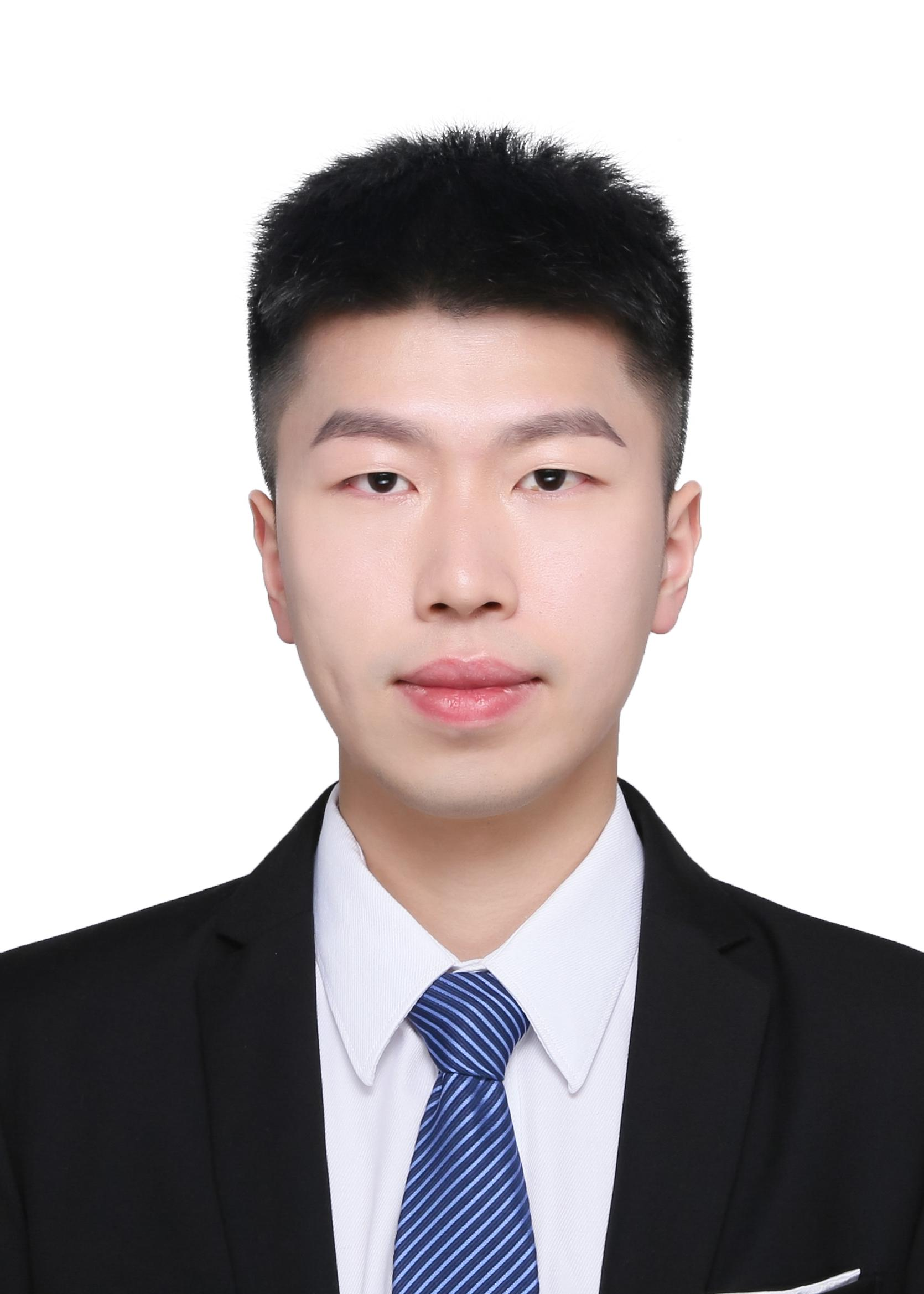}}]{Xiang Zhu}
		(Graduate Student Member, IEEE) received his B.E. degree in automation from University of Electronic Science and Technology of China, Chengdu, China in 2022. He is currently pursuing his Ph.D. degree in control science and engineering at Department of Automation, Tsinghua University. His research interests include the optimization of distributed energy resources, ancillary services of virtual power plants and fast frequency response.
	\end{IEEEbiography}

	\begin{IEEEbiography}[{\includegraphics[width=1in,clip,keepaspectratio]{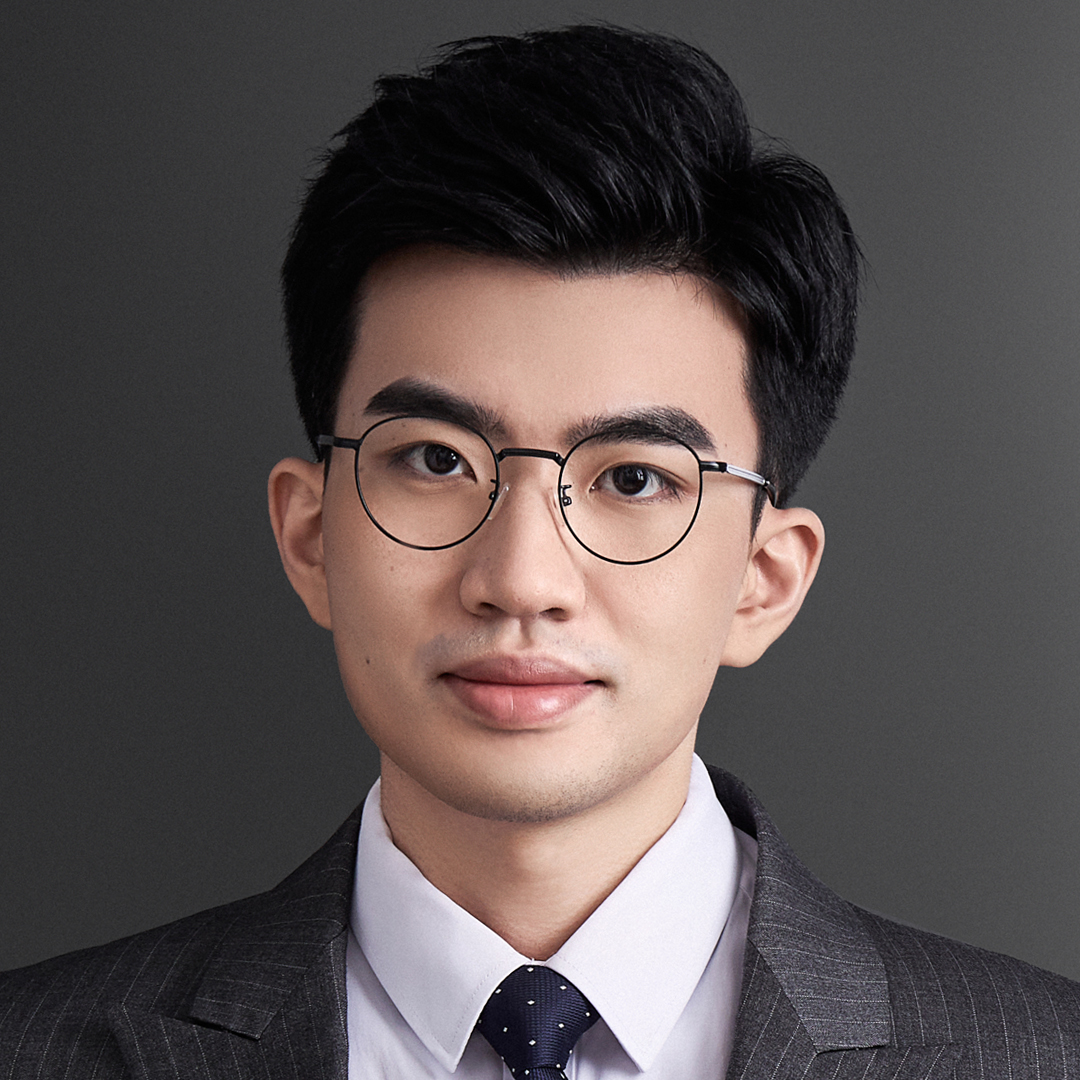}}]{Guangchun (Grant) Ruan}
		(Member, IEEE) is currently a postdoc with the Laboratory for Information \& Decision Systems (LIDS) at MIT. Before joining MIT, he received the Ph.D. degree in electrical engineering from Tsinghua University in 2021, and worked as a postdoc with the University of Hong Kong in 2022. He visited Texas A\&M University in 2020 and the University of Washington in 2019. His research interests include electricity market, energy resilience, demand response, data science and machine learning applications.
	\end{IEEEbiography}

	\begin{IEEEbiography}[{\includegraphics[width=1in,clip,keepaspectratio]{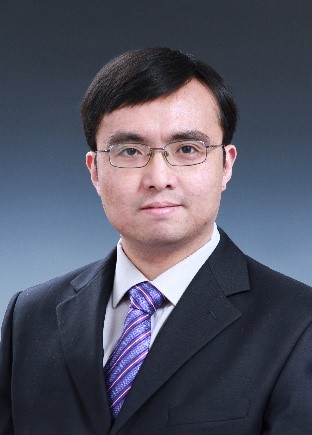}}]{Hua Geng}
		(Fellow, IEEE) received the B.S. degree in electrical engineering from Huazhong University of Science and Technology, Wuhan, China, in 2003 and the Ph.D. degree in control theory and application from Tsinghua University, Beijing, China, in 2008. From 2008 to 2010, he was a Postdoctoral Research Fellow with the Department of Electrical and Computer Engineering, Ryerson University, Toronto, ON, Canada. He joined Automation Department of Tsinghua University in June 2010 and is currently a full professor.
		
		His current research interests include advanced control on power electronics and renewable energy conversion systems, AI for energy systems. He has authored more than 300 technical publications and holds more than 30 issued Chinese/US patents. He was the recipient of IEEE PELS Sustainable Energy Systems Technical Achievement Award. He is the Editor-in-Chief of IEEE Transactions on Sustainable Energy. He served as general chair, track chairs and session chairs of several IEEE conferences. He is an IEEE Fellow and an IET Fellow, convener of the modeling working group in IEC SC 8A.
	\end{IEEEbiography}

	\begin{IEEEbiography}[{\includegraphics[width=1in,clip,keepaspectratio]{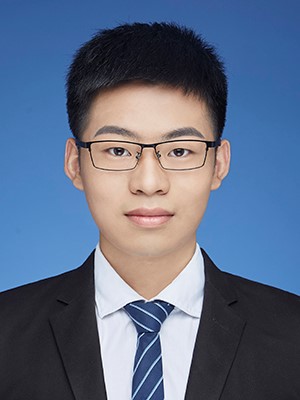}}]{Honghai Liu}
		received his B.E. degree in electrical engineering from North China Electric Power University, Beijing, China, in 2022. He is currently pursuing his Ph.D. degree in electrical engineering at the North China Electric Power University, Beijing, China. His research interests include optimization of virtual power plant and deep reinforcement learning applications in power systems.
	\end{IEEEbiography}

	\begin{IEEEbiography}[{\includegraphics[width=1in,clip,keepaspectratio]{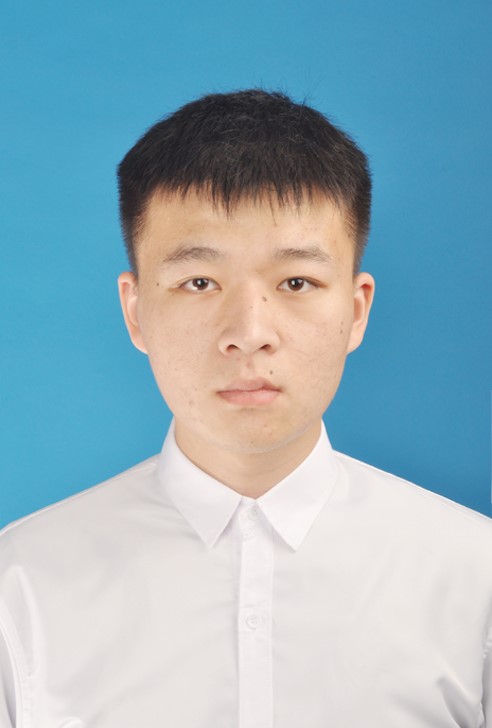}}]{Minfei Bai}
		received his B.E. degree in electrical engineering from Tianjin University of Technology, Tianjin, China, in 2022. He is currently pursuing his M. degree in electrical engineering at the North China Electric Power University, Beijing, China. His research interests include robust optimization and its applications in power system scheduling and planning.
	\end{IEEEbiography}
	
	\begin{IEEEbiography}[{\includegraphics[width=1in,clip,keepaspectratio]{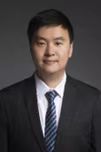}}]{Chao Peng}
		received his M.S. and Ph.D. in Automation from the University of Electronic Science and Technology of China in 2007 and 2012, respectively, and is currently an associate professor at the School of Automation Engineering at the University of Electronic Science and Technology of China. His research interests include renewable power generation system modeling and control, multi-energy system integration and operation optimization.
	\end{IEEEbiography}

\end{document}